# A recursive field-normalized bibliometric performance indicator: An application to the field of library and information science

Ludo Waltman,[1] Erjia Yan,[2] and Nees Jan van Eck[1]

[1] Centre for Science and Technology Studies, Leiden University, Leiden, The Netherlands
{waltmanlr, ecknjpvan}@cwts.leidenuniv.nl

[2] School of Library and Information Science, Indiana University, Bloomington, Indiana, United States
eyan@indiana.edu

Two commonly used ideas in the development of citation-based research performance indicators are the idea of normalizing citation counts based on a field classification scheme and the idea of recursive citation weighing (like in PageRank-inspired indicators). We combine these two ideas in a single indicator, referred to as the recursive mean normalized citation score indicator, and we study the validity of this indicator. Our empirical analysis shows that the proposed indicator is highly sensitive to the field classification scheme that is used. The indicator also has a strong tendency to reinforce biases caused by the classification scheme. Based on these observations, we advise against the use of indicators in which the idea of normalization based on a field classification scheme and the idea of recursive citation weighing are combined.

## 1. Introduction

In bibliometric and scientometric research, there is a trend towards developing more and more sophisticated citation-based research performance indicators. In this paper, we are concerned with two streams of research. One stream of research focuses on the development of indicators that aim to correct for the fact that the density of citations (i.e., the average number of citations per publication) differs among fields. Two basic approaches can be distinguished. One approach is to normalize citation counts for field differences based on a classification scheme that assigns publications to fields (e.g., Braun & Glänzel, 1990; Moed, De Bruin, & Van Leeuwen, 1995; Waltman, Van Eck, Van Leeuwen, Visser, & Van Raan, 2011). The other approach is to normalize citation counts based on the number of references in citing publications or citing journals (e.g., Moed, 2010; Zitt & Small, 2008). The latter approach, which is sometimes referred to as source normalization (Moed, 2010), does not need a field classification scheme.

A second stream of research focuses on the development of recursive indicators, typically inspired by the well-known PageRank algorithm (Brin & Page, 1998). In the case of recursive indicators, citations are weighed differently depending on the status of the citing publication (e.g., Chen, Xie, Maslov, & Redner, 2007; Ma, Guan, & Zhao, 2008; Walker, Xie, Yan, & Maslov, 2007), the citing journal (e.g., Bollen, Rodriguez, & Van de Sompel, 2006; Pinski & Narin, 1976), or the citing author (e.g., Radicchi et al., 2009; Życzkowski, 2010). The underlying idea is that a citation from an influential publication, a prestigious journal, or a renowned author should be regarded as more valuable than a citation from an insignificant publication, an obscure journal, or an unknown author. It is sometimes argued that non-recursive indicators measure popularity while recursive indicators measure prestige (e.g., Bollen et al., 2006; Yan & Ding, 2010).

Based on the above discussion, we have two binary dimensions along which we can distinguish citation-based research performance indicators, namely the dimension of normalization based on a field classification scheme versus source normalization and the dimension of non-recursive mechanisms versus recursive mechanisms. These two dimensions yield four types of indicators. This is shown in Table 1, in which we list some examples of the different types of indicators. It is important to note that all currently existing indicators that use a classification scheme for normalizing citation counts are of a non-recursive nature. Hence, there currently are no recursive indicators that make use of a classification scheme.[1] Instead, the currently existing recursive indicators can best be regarded as belonging to the family of source-normalized indicators. This is because these indicators, like non-recursive source-normalized indicators, are based in one way or another on the idea that each unit (i.e., each publication, journal, or author) has a certain weight which it distributes over the units it cites. We refer to Waltman and Van Eck (2010a) for a detailed analysis of the close relationship between a source-normalized indicator (i.e., the audience factor) and two recursive indicators (i.e., the Eigenfactor indicator and the influence weight indicator).

Table 1. A classification of some citation-based research performance indicators based on their normalization approach and the presence or absence of a recursive mechanism.

|  | Normalization based on classification scheme | Source normalization |
|---|---|---|
| Non-recursive mechanism | Citation $z$-score (Lundberg, 2007)<br>CPP/FCSm (Moed et al., 1995)<br>MNCS (Waltman et al., 2011)<br>NMCR (Braun & Glänzel, 1990) | Audience factor (Zitt, 2010; Zitt & Small, 2008)<br>Fractional counting (Glänzel et al., 2011; Leydesdorff & Bornmann, 2011)<br>SNIP (Moed, 2010)<br>Source-normalized MNCS (Waltman & Van Eck, 2010b) |
| Recursive mechanism |  | CiteRank (Walker et al., 2007)<br>Eigenfactor (Bergstrom, 2007; West et al., 2010)<br>Influence weight (Pinski & Narin, 1976)<br>Science author rank (Radicchi et al., 2009)<br>SCImago journal rank (González-Pereira et al., 2010)<br>Weighted PageRank (Bollen et al., 2006) |

In this paper, we focus on the empty cell in the lower left of Table 1. Hence, we focus on recursive indicators that use a field classification scheme for normalizing citation counts. We first propose a recursive variant of the mean normalized citation score (MNCS) indicator (Waltman et al., 2011). We then present an empirical analysis of this recursive MNCS indicator. In the analysis, the recursive MNCS indicator is used to study the citation impact of journals and research institutes in the field of library and information science. Our aim is to get insight into the validity of recursive indicators that use a classification scheme for normalizing citation counts. We pay special attention to the sensitivity of such indicators to the classification scheme that is used.

---
[1] However, a first step in the direction of such indicators was taken by Van Leeuwen, Visser, Moed, Nederhof, and Van Raan (2003). They proposed an indicator that weighs citations by the average field-normalized number of citations per publication of the citing journal.

## 2. Recursive mean normalized citation score

The ordinary non-recursive MNCS indicator for a set of publications equals the average number of citations per publication, where for each publication the number of citations is normalized for differences among fields (Waltman et al., 2011). The normalization is performed by dividing the number of citations of a publication by the publication's expected number of citations. The expected number of citations of a publication is defined as the average number of citations per publication in the field in which the publication was published. An example of the calculation of the non-recursive MNCS indicator is provided in Table 2.

Table 2. Example of the calculation of the ordinary non-recursive MNCS indicator. There are three publications. For each publication, the table lists the number of citations, the field, the expected number of citations, and the normalized citation score. The normalized citation score of a publication is obtained by dividing the number of citations by the expected number of citations. The MNCS indicator equals the average of the normalized citation scores of the three publications.

| Publication | No. cit. | Field | Expected no. cit. | Normalized cit. score |
|---|---|---|---|---|
| A | 3 | X | 4.32 | 0.69 |
| B | 8 | X | 4.32 | 1.85 |
| C | 10 | Y | 12.17 | 0.82 |

MNCS = (0.69 + 1.85 + 0.82) / 3 = 1.12

The non-recursive MNCS indicator can also be referred to as the first-order MNCS indicator. We define the second-order MNCS indicator in the same way as the first-order MNCS indicator except that citations are weighed differently. In the first-order MNCS indicator, all citations have the same weight. In the second-order MNCS indicator, on the other hand, the weight of a citation is given by the value of the first-order MNCS indicator for the citing journal. Hence, citations from journals with a high value for the first-order MNCS indicator are regarded as more valuable than citations from journals with a low value for the first-order MNCS indicator. We have now defined the second-order MNCS indicator in terms of the first-order MNCS indicator. In the same way, we define the third-order MNCS indicator in terms of the second-order MNCS indicator, the fourth-order MNCS indicator in terms of the third-order MNCS indicator, and so on. This yields the recursive MNCS indicator that we study in this paper.

To make the above definition of the recursive MNCS indicator more precise, we formalize it mathematically. We use $i$, $j$, $k$, and $l$ to denote, respectively, a publication, a journal, a field, and an institute. We define

$$c_{ii'} = \begin{cases} 1 & \text{if publication } i \text{ cites publication } i' \\ 0 & \text{otherwise,} \end{cases} \quad (1)$$

$$p_{ij} = \begin{cases} 1 & \text{if publication } i \text{ is published in journal } j \\ 0 & \text{otherwise,} \end{cases} \quad (2)$$

$$b_{ik} = \begin{cases} 1 & \text{if publication } i \text{ belongs to field } k \\ 0 & \text{otherwise,} \end{cases} \quad (3)$$

$$a_{il} = \begin{cases} 1 & \text{if publication } i \text{ is authored by institute } l \\ 0 & \text{otherwise.} \end{cases} \quad (4)$$

For $\alpha = 1, 2, \ldots$, the $\alpha$th-order citation score of publication $i$ is defined as

$$\text{CS}_i^{(\alpha)} = \sum_{i'} w_{i'}^{(\alpha)} c_{i'i}, \quad (5)$$

where $w_{i'}^{(\alpha)}$ denotes the weight of a citation from publication $i'$. For $\alpha = 1$, $w_{i'}^{(\alpha)} = 1$ for all $i'$. For $\alpha = 2, 3, \ldots$, $w_{i'}^{(\alpha)}$ is given by

$$w_{i'}^{(\alpha)} = \sum_j p_{i'j} \text{MNCS}_j^{(\alpha-1)}. \quad (6)$$

It follows from (5) and (6) that the $\alpha$th-order citation score of a publication equals a weighted sum of the citations received by the publication. For $\alpha = 1$, all citations have the same weight. For $\alpha = 2, 3, \ldots$, the weight of a citation is given by the $(\alpha - 1)$th-order MNCS of the citing journal.

We define the $\alpha$th-order mean citation score of a field as the average $\alpha$th-order citation score of all publications belonging to the field, that is,

$$\text{MCS}_k^{(\alpha)} = \frac{\sum_i b_{ik} \text{CS}_i^{(\alpha)}}{\sum_i b_{ik}}. \quad (7)$$

The $\alpha$th-order expected citation score of a publication is defined as the $\alpha$th-order mean citation score of the field to which the publication belongs,[2] and the $\alpha$th-order normalized citation score of a publication is defined as the ratio of the publication's $\alpha$th-order citation score and its $\alpha$th-order expected citation score. This yields

$$\text{ECS}_i^{(\alpha)} = \sum_k b_{ik} \text{MCS}_k^{(\alpha)}, \quad (8)$$

$$\text{NCS}_i^{(\alpha)} = \frac{\text{CS}_i^{(\alpha)}}{\text{ECS}_i^{(\alpha)}}. \quad (9)$$

---

[2] For simplicity, we assume that fields are non-overlapping. A publication therefore always belongs to exactly one field.

If the $\alpha$th-order normalized citation score of a publication is greater (less) than one, this indicates that the $\alpha$th-order citation score of the publication is greater (less) than the average $\alpha$th-order citation score of all publications in the field.

We define the $\alpha$th-order MNCS of a set of publications as the average $\alpha$th-order normalized citation score of the publications in the set. In the case of journals and institutes, we obtain respectively

$$\text{MNCS}_j^{(\alpha)} = \frac{\sum_i p_{ij} \text{NCS}_i^{(\alpha)}}{\sum_i p_{ij}}, \qquad (10)$$

$$\text{MNCS}_l^{(\alpha)} = \frac{\sum_i \bar{a}_{il} \text{NCS}_i^{(\alpha)}}{\sum_i \bar{a}_{il}}, \qquad (11)$$

where

$$\bar{a}_{il} = \frac{a_{il}}{\sum_{l'} a_{il'}}. \qquad (12)$$

It follows from (11) and (12) that in the case of institutes we take a fractional counting approach. That is, a publication resulting from a collaboration of, say, three institutes is counted for each institute as 1/3 of a full publication. Alternatively, a full counting approach could have been taken. A collaborative publication would then be counted as a full publication for each of the institutes involved. A full counting approach is obtained by replacing $\bar{a}_{il}$ by $a_{il}$ in (11).

Until now, we have only discussed the issue of normalization for the field in which a publication was published. We have not discussed the issue of normalization for the age of a publication. The latter type of normalization can be used to correct for the fact that older publications have had more time to earn citations than younger publications. Normalization for the age of a publication can easily be incorporated into indicators that make use of a field classification scheme, such as the MNCS indicator. It is more difficult to incorporate into (recursive or non-recursive) source-normalized indicators (see however Waltman & Van Eck, 2010b). In the empirical analysis presented in Section 4, the recursive MNCS indicator performs a normalization both for the field in which a publication was published and for the age of a publication. This means that in the above mathematical description of the recursive MNCS indicator $k$ in fact represents not just a field but a combination of a field and a publication year. As a consequence, $b_{ik}$ indicates whether a publication was published in a certain field and year, and $\text{MCS}_k^{(\alpha)}$ indicates the average $\alpha$th-order citation score of all publications published in a certain field and year.

## 3. Data

To test our recursive MNCS indicator, we use the indicator to study the citation impact of journals and research institutes in the field of library and information science (LIS). We focus on the period from 2000 to 2009. Our analysis is based on data from the Web of Science database.

We first needed to delineate the LIS field. We used the *Journal of the American Society for Information Science and Technology* (*JASIST*) as the 'seed' journal for our delineation. We decided to select the 47 journals that, based on co-citation data, are most strongly related with *JASIST*. Only journals in the Web of Science subject category *Information Science & Library Science* were considered. *JASIST* together with the 47 selected journals constituted our delineation of the LIS field. From the journals within our delineation, we selected all 12,202 publications in the period 2000–2009 that are of the document type 'article' or 'review'. It is important to emphasize that in our analysis we only take into account citations within the set of 12,202 publications. Citations given by publications outside this set are not considered. Self citations are also excluded.

Table 3. 48 LIS journals and their assignment to three clusters.

| *Library science (27 journals)* | *Information science (14 journals)* |
|---|---|
| African Journal of Library Archives and Information Science | Annual Review of Information Science and Technology |
| Australian Library Journal | Aslib Proceedings |
| College & Research Libraries | Canadian Journal of Information and Library Science |
| Electronic Library | |
| Information Technology and Libraries | Information Processing & Management |
| Interlending & Document Supply | Information Research |
| International Information & Library Review | Journal of Documentation |
| Journal of Academic Librarianship | Journal of Information Science |
| Journal of Librarianship and Information Science | Journal of the American Society for Information Science |
| Journal of Scholarly Publishing | |
| Law Library Journal | Journal of the American Society for Information Science and Technology |
| Learned Publishing | |
| Library & Information Science Research | Knowledge Organization |
| Library and Information Science | Online Information Review |
| Library Collections Acquisitions & Technical Services | Perspectivas Em Ciencia Da Informacao |
| | Proceedings of the ASIST Annual Meeting |
| Library Hi Tech | Profesional De La Informacion |
| Library Quarterly | |
| Library Resources & Technical Services | *Scientometrics (7 journals)* |
| Library Trends | ASIST Monograph Series |
| Libri | Australian Academic & Research Libraries |
| Malaysian Journal of Library & Information Science | Investigacion Bibliotecologica |
| | Journal of Informetrics |
| Portal-Libraries and the Academy | Research Evaluation |
| Program-Electronic Library and Information Systems | Revista Espanola De Documentacion Cientifica |
| | Scientometrics |
| Reference & User Services Quarterly | |
| Science & Technology Libraries | |
| Serials Librarian | |
| Serials Review | |

In our analysis, we also study the effect of splitting up the LIS field in a number of subfields. We used the following procedure to split up the LIS field. We first collected bibliographic coupling data for the 48 journals in our analysis.[3] Based on the bibliographic coupling data, we created a clustering of the journals. We used the VOS clustering technique (Waltman, Van Eck, & Noyons, 2010), available in the VOSviewer software (Van Eck & Waltman, 2010), to create the clustering. After some experimenting, we decided that the most satisfactory clustering has three clusters. These clusters can roughly be interpreted as follows. The largest cluster (27 journals) deals with library science, the smallest cluster (7 journals) deals with scientometrics, and the third cluster (14 journals) deals with general information science topics. The assignment of the 48 journals to the three clusters is shown in Table 3. The clustering of the journals is also shown in the journal map in Figure 1. This map, produced using the VOSviewer software, is based on bibliographic coupling relations between the journals.

---
[3] We also tried out using co-citation data, but this gave less satisfactory results.

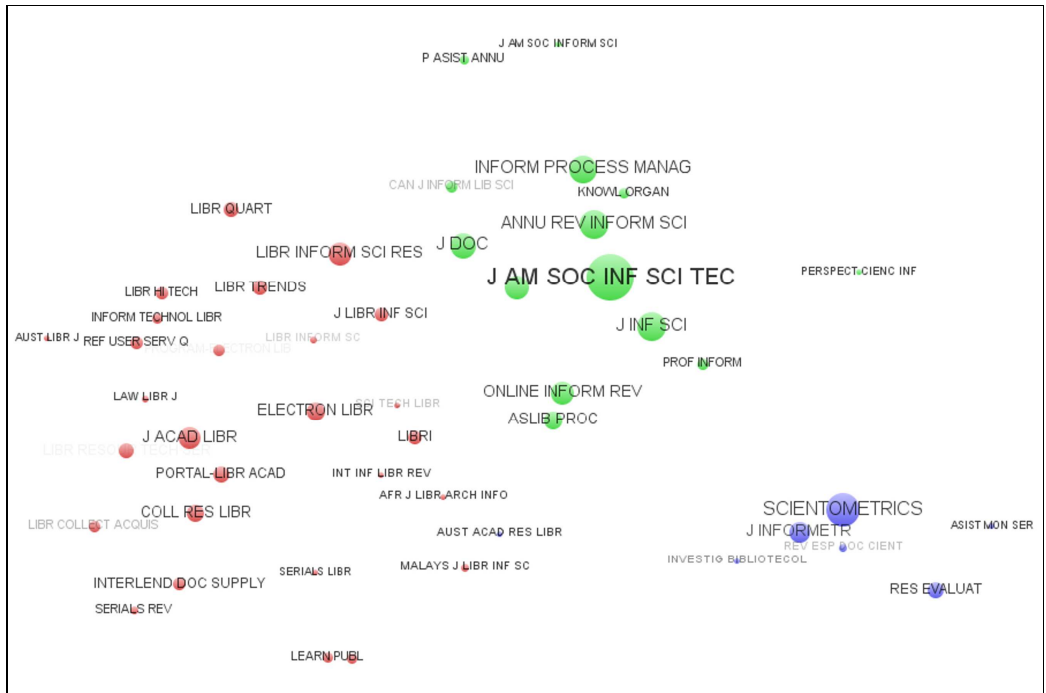

Figure 1. Journal map of 48 LIS journals based on bibliographic coupling data. The color of a journal indicates the cluster to which it belongs. The map was produced using the VOSviewer software.

## 4. Results

As discussed in the previous section, we can treat LIS either as a single integrated field or as a field consisting of three separate subfields (i.e., library science, information science, and scientometrics). In the latter case, the recursive MNCS indicator normalizes for differences among the three subfields in the average number of citations per publication. Below, we first present the results obtained when LIS is treated as a single integrated field. We then present the results obtained when LIS is treated as a field consisting of three separate subfields. We also present a comparison of the results obtained using the two approaches. We emphasize once more that in our analysis citations given by publications outside our set of 12,202 publications are not taken into account.

### 4.1. Single integrated LIS field

We first consider the case of a single integrated LIS field. The recursive MNCS indicator is said to have converged for a certain $\alpha$ if there is virtually no difference between values of the $\alpha$th-order MNCS indicator and values of the $(\alpha + 1)$th-order MNCS indicator. For our data, convergence of the recursive MNCS indicator can be observed for $\alpha = 20$. In our analysis, our main focus therefore is on comparing the first-order MNCS indicator (i.e., the ordinary non-recursive MNCS indicator) with the 20th-order MNCS indicator.

In Table 4, we list the top 10 journals according to both the first-order MNCS indicator and the 20th-order MNCS indicator. In the case of the first-order MNCS indicator, the top 10 consists of journals from all three subfields. However, journals from the information science and scientometrics subfields seem to slightly dominate journals from the library science subfield. There are three library science journals in the top 10, at ranks 4, 8, and 10. Given that more than half of the journals in our analysis belong to the library science subfield (27 of the 48 journals), the library science journals seem to be underrepresented in the top 10. Also, within the top 10, the three library science journals have relatively low ranks.

Table 4. Top 10 journals according to both the first-order MNCS indicator and the 20th-order MNCS indicator. LIS is treated as a single integrated field in the calculation of the indicators.

| Journal | MNCS ($\alpha = 1$) | Journal | MNCS ($\alpha = 20$) |
|---|---|---|---|
| Journal of Informetrics | 4.49 | Journal of Informetrics | 12.32 |
| Annual Review of Information Science and Technology | 2.97 | Annual Review of Information Science and Technology | 3.79 |
| Journal of the American Society for Information Science | 2.35 | Scientometrics | 3.17 |
| Interlending & Document Supply | 1.94 | Journal of the American Society for Information Science | 2.72 |
| Journal of the American Society for Information Science and Technology | 1.84 | Journal of the American Society for Information Science and Technology | 2.36 |
| Scientometrics | 1.72 | Journal of Documentation | 1.36 |
| Journal of Documentation | 1.58 | Information Processing & Management | 1.21 |
| College & Research Libraries | 1.33 | Journal of Information Science | 0.96 |
| Information Processing & Management | 1.21 | Library & Information Science Research | 0.82 |
| Library & Information Science Research | 1.17 | Research Evaluation | 0.81 |

Let's now turn to the top 10 journals according to the 20th-order MNCS indicator. This top 10 provides a much more extreme picture. The top 10 is now almost completely dominated by information science and scientometrics journals. There is only one library science journal left, at rank 9. Moreover, when looking at the values of the MNCS indicator, large differences can be observed within the top 10. Especially the extremely high value of the MNCS indicator for *Journal of Informetrics*, the highest ranked journal, is striking. The value of the MNCS indicator for this journal is more than three times as high as the value of the MNCS indicator for *Annual Review of Information Science and Technology*, which is the second-highest ranked journal.

In Figure 2, the first-order MNCS indicator is compared with the second-order MNCS indicator (left panel) and the 20th-order MNCS indicator (right panel) for the 48 LIS journals in our data set. As can be seen in the figure, the differences between the first- and the second-order MNCS indicator are relatively small, although there is one journal (*Journal of Informetrics*) that benefits quite a lot from going from the first-order MNCS indicator to the second-order MNCS indicator. The differences between the first- and the 20th-order MNCS indicator are much larger. As was also seen in Table 4, there are a number of journals for which the value of the 20th-order MNCS indicator is higher than the value of the first-order MNCS indicator. These are all information science and scientometrics journals. The library science journals all turn out to have a lower value for the 20th-order MNCS indicator than for the first-order MNCS indicator.

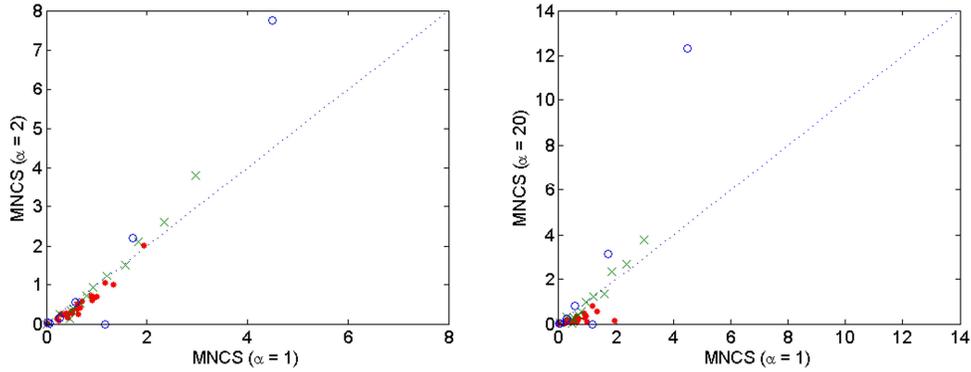

Figure 2. Comparison of the first-order MNCS indicator with the second-order MNCS indicator (left panel) and the 20th-order MNCS indicator (right panel) for 48 LIS journals. LIS is treated as a single integrated field in the calculation of the indicators. Library science, information science, and scientometrics journals are indicated by, respectively, red points, green crosses, and blue circles.

Table 5. Top 10 research institutes according to both the first-order MNCS indicator and the 20th-order MNCS indicator. LIS is treated as a single integrated field in the calculation of the indicators.

| Institute | MNCS ($\alpha = 1$) | Institute | MNCS ($\alpha = 20$) |
|---|---|---|---|
| Leiden Univ | 3.85 | Univ Antwerp | 8.64 |
| Univ Antwerp | 3.77 | Hungarian Acad Sci | 7.92 |
| Hungarian Acad Sci | 3.70 | Univ Amsterdam | 7.45 |
| Univ Amsterdam | 3.53 | Leiden Univ | 7.43 |
| Royal Sch Lib & Informat Sci | 3.20 | Limburg Univ Ctr | 5.94 |
| Indiana Univ | 2.52 | Kathol Univ Leuven | 5.19 |
| Hebrew Univ Jerusalem | 2.51 | Indiana Univ | 3.56 |
| Kathol Univ Leuven | 2.48 | Univ Wolverhampton | 3.13 |
| Univ Tennessee - Knoxville | 2.39 | Royal Sch Lib & Informat Sci | 3.07 |
| Univ Bar Ilan | 2.35 | Hebrew Univ Jerusalem | 2.72 |

In addition to journals, we also consider research institutes in LIS. We restrict our analysis to the 86 institutes that have at least 25 publications in our data set. (Recall from Section 2 that publications are counted fractionally.) The top 10 institutes according to both the first-order MNCS indicator and the 20th-order MNCS indicator are listed in Table 5. Comparing the results of the two MNCS indicators, it is clear that institutes which are mainly active in the scientometrics subfield benefit a lot from the use of a higher-order MNCS indicator. For these institutes, the value of the 20th-order MNCS indicator tends to be much higher than the value of the first-order MNCS indicator. This is consistent with our above analysis for LIS journals, where we found that the two most important scientometrics journals (*Journal of Informetrics* and *Scientometrics*) benefit quite significantly from the use of a higher-order MNCS indicator.

**4.2. Three separate LIS subfields**

We now consider the case in which LIS is divided into three separate subfields (i.e., library science, information science, and scientometrics). In the calculation of the recursive MNCS indicator, a normalization is performed to correct for differences among the three subfields in the average number of citations per publication. Like in the above analysis, we focus mainly on comparing the first-order MNCS indicator with the 20th-order MNCS indicator.

In Table 6, the top 10 journals according to both the first-order MNCS indicator and the 20th-order MNCS indicator is shown. This table is similar to Table 4 above, except that in the calculation of the indicators LIS is treated as a field consisting of three separate subfields rather than as a single integrated field. Comparing Table 6 with Table 4, it can be seen that library science journals now play a much more prominent role, both in the case of the first-order MNCS indicator and in the case of the 20th-order MNCS indicator. As a consequence, the top 10 journals now looks much more balanced for both MNCS indicators. Also, unlike in Table 4, there are no journals in Table 6 with an extremely high value for the 20th-order MNCS indicator.

Table 6. Top 10 journals according to both the first-order MNCS indicator and the 20th-order MNCS indicator. LIS is treated as a field consisting of three separate subfields in the calculation of the indicators.

| Journal | MNCS ($\alpha = 1$) | Journal | MNCS ($\alpha = 20$) |
|---|---|---|---|
| Journal of Informetrics | 3.07 | Interlending & Document Supply | 5.25 |
| Annual Review of Information Science and Technology | 2.62 | Journal of Informetrics | 4.20 |
| Interlending & Document Supply | 2.46 | Annual Review of Information Science and Technology | 3.40 |
| College & Research Libraries | 1.90 | College & Research Libraries | 1.83 |
| Library & Information Science Research | 1.67 | Journal of the American Society for Information Science and Technology | 1.81 |
| Journal of the American Society for Information Science and Technology | 1.66 | Library & Information Science Research | 1.75 |
| Journal of the American Society for Information Science | 1.44 | Serials Review | 1.52 |
| Serials Review | 1.39 | Journal of the American Society for Information Science | 1.52 |
| Portal-Libraries and the Academy | 1.36 | Learned Publishing | 1.37 |
| Journal of Documentation | 1.35 | Journal of Documentation | 1.26 |

As can be seen in Table 6, the journal with the highest value for the 20th-order MNCS indicator is *Interlending & Document Supply*. We investigated this journal in more detail and found that each issue of the journal contains a review article entitled "Interlending and document supply: A review of the recent literature". These review articles refer to other articles published in the same issue of the journal. Clearly, the practice of publishing these review articles is an important contributing factor to the journal's top ranking in Table 6.

In Figure 3, a comparison is presented of the first-order MNCS indicator with the second-order MNCS indicator (left panel) and the 20th-order MNCS indicator (right panel) for the 48 LIS journals in our data set. The figure shows that most of the differences between the first-order MNCS indicator and the higher-order MNCS indicators are not very large, especially when compared with the differences shown in Figure 2. The journal that is most sensitive to the use of a higher-order MNCS indicator is *Interlending & Document Supply*. As pointed out above, this journal has quite special citation characteristics, which explains its sensitivity to the use of a higher-order MNCS indicator.

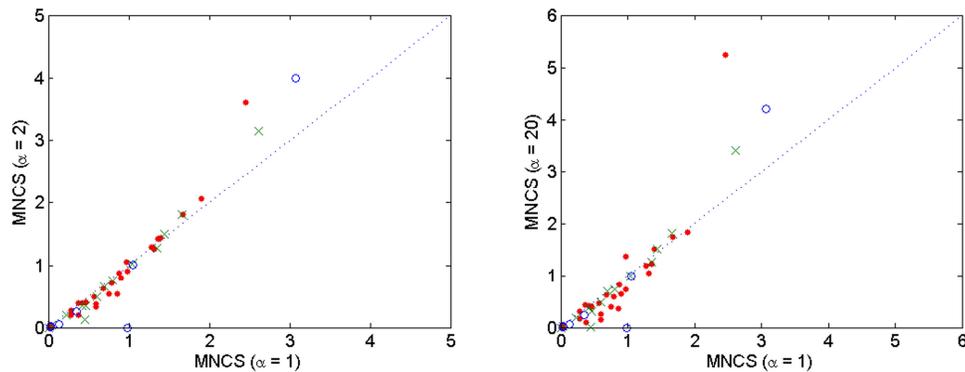

Figure 3. Comparison of the first-order MNCS indicator with the second-order MNCS indicator (left panel) and the 20th-order MNCS indicator (right panel) for 48 LIS journals. LIS is treated as a field consisting of three separate subfields in the calculation of the indicators. Library science, information science, and scientometrics journals are indicated by, respectively, red points, green crosses, and blue circles.

Table 7. Top 10 research institutes according to both the first-order MNCS indicator and the 20th-order MNCS indicator. LIS is treated as a field consisting of three separate subfields in the calculation of the indicators.

| Institute | MNCS ($\alpha = 1$) | Institute | MNCS ($\alpha = 20$) |
|---|---|---|---|
| Univ Amsterdam | 3.04 | Univ Amsterdam | 3.64 |
| Univ Antwerp | 2.79 | Univ Antwerp | 3.63 |
| Royal Sch Lib & Informat Sci | 2.67 | Limburg Univ Ctr | 2.84 |
| Leiden Univ | 2.60 | Leiden Univ | 2.74 |
| Hungarian Acad Sci | 2.44 | Hungarian Acad Sci | 2.58 |
| Cornell Univ | 2.34 | Indiana Univ | 2.54 |
| Indiana Univ | 2.26 | Univ Tennessee - Knoxville | 2.46 |
| Univ Tennessee - Knoxville | 2.23 | Univ Coll Dublin | 2.40 |
| Hebrew Univ Jerusalem | 2.02 | Cornell Univ | 2.34 |
| Univ Wolverhampton | 1.97 | Royal Sch Lib & Informat Sci | 2.32 |

Results for research institutes in LIS are reported in Table 7. In the table, the top 10 institutes according to both the first-order MNCS indicator and the 20th-order MNCS indicator are shown. Like in Table 5, the top of the ranking is dominated by institutes with a strong focus on scientometrics research. This is the case both for the first-order MNCS indicator and for the 20th-order MNCS indicator. However, comparing Table 7 with Table 5, it can be seen that the MNCS values of the scientometrics institutes have decreased quite considerably, especially when looking at the 20th-order MNCS indicator. Hence, although the scientometrics institutes still occupy the top positions in the ranking, the differences with the other institutes have become smaller.

**4.3. Comparison**

In Figure 4, we present a direct comparison of on the one hand the results obtained when LIS is treated as a single integrated field and on the other hand the results obtained when LIS is treated as a field consisting of three separate subfields. The comparison is made for the 48 LIS journals in our data set. Figure 4 clearly shows how treating LIS as a single integrated field benefits the scientometrics journals and harms the library science journals. In addition, the figure shows that this effect is strongly reinforced when instead of the first-order MNCS indicator the 20th-order MNCS indicator is used.

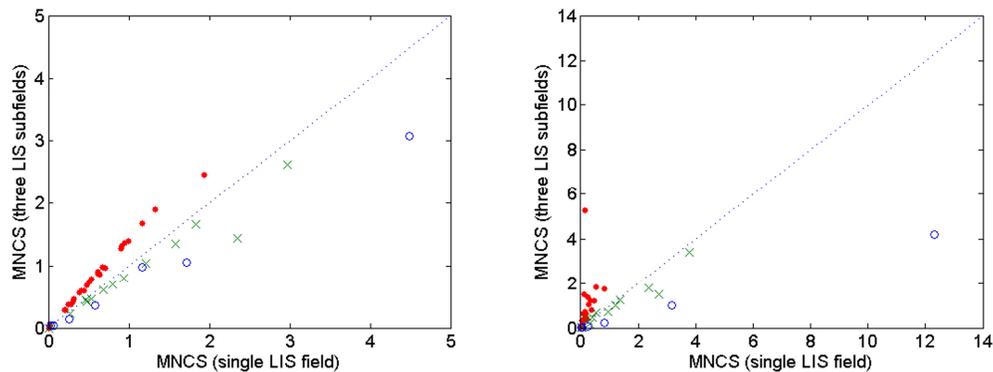

Figure 4. Comparison of the MNCS indicator when LIS is treated as a single integrated field with the MNCS indicator when LIS is treated as a field consisting of three separate subfields. The comparison is made for 48 LIS journals, and either the first-order MNCS indicator (left panel) or the 20th-order MNCS indicator (right panel) is used. Library science, information science, and scientometrics journals are indicated by, respectively, red points, green crosses, and blue circles.

## 5. Discussion and conclusion

Recursive bibliometric indicators are based on the idea that citations should be weighed differently depending on the source from which they originate. Citations from a prestigious journal, for instance, should have more weight than citations from an obscure journal. It is sometimes argued that by weighing citations differently depending on their source it is possible to measure not just the popularity of publications but also their prestige.

In this paper, we have combined the idea of recursive citation weighing with the idea of using a classification scheme to normalize citation counts for differences among fields. The combination of these two ideas has not been explored before. Although when used separately from each other the two ideas can be quite useful, our empirical analysis for the field of LIS indicates that the combination of the two ideas does not yield satisfactory results. The main observations from our analysis are twofold. First, our proposed recursive MNCS indicator is highly sensitive to the way in which fields are defined in the classification scheme that one uses. And second, if within a field there are subfields with significantly different citation characteristics, the recursive MNCS indicator will be strongly biased in favor of the subfields with the highest density of citations.

The sensitivity of bibliometric indicators to the field classification scheme that is used for normalizing citation counts has been investigated in various studies (Adams, Gurney, & Jackson, 2008; Bornmann, Mutz, Neuhaus, & Daniel, 2008; Neuhaus, & Daniel, 2009; Van Leeuwen, & Calero Medina, 2009; Zitt, Ramanana-Rahary, & Bassecoulard, 2005). Our empirical results for the first-order MNCS indicator (i.e., the ordinary non-recursive MNCS indicator) are in line with earlier studies in which a significant sensitivity of bibliometric indicators to the classification scheme that is used has been reported. For the 20th-order MNCS indicator, this sensitivity even turns out to be much higher. Treating LIS as a single integrated field or as a field consisting of three separate subfields yields very different results for the 20th-order MNCS indicator.

If a field as defined in the classification scheme that one uses is heterogeneous in terms of citation characteristics, bibliometric indicators will have a bias that favors subfields with a higher citation density over subfields with a lower citation density. This is a general problem of bibliometric indicators that use a classification scheme to normalize citation counts. In the case of the recursive MNCS indicator, our empirical results show that the idea of recursive citation weighing strongly reinforces biases caused by the classification scheme. Within the field of LIS, the scientometrics subfield has the highest citation density, followed by the information science subfield. The library science subfield has the lowest citation density. When LIS is treated as a single integrated field, the differences in citation density among the three LIS subfields cause both the first-order MNCS indicator and the 20th-order MNCS indicator to be biased, where the bias favors the scientometrics subfield and harms the library science subfield. However, the bias is much stronger for the 20th-order MNCS indicator than for the first-order MNCS indicator. For instance, in the case of the 20th-order MNCS indicator, library science journals are completely dominated by journals in scientometrics and information science.

Based on the above observations, we advise against the introduction of recursiveness into bibliometric indicators that use a field classification scheme for normalizing citation counts (such as the MNCS indicator). Instead of providing more sophisticated measurements of citation impact (measurements of 'prestige' rather than 'popularity'), the main effect of introducing recursiveness is to reinforce biases caused by the way in which fields are defined in the classification scheme that one uses.